\documentclass{optica-article}

\journal{opticajournal} 

\articletype{Research Article}

\usepackage{subcaption}
\usepackage{booktabs}

\begin{document}

\title{Simulation-Driven Approach for Occlusion-Based Non-Line-Of-Sight Detection Using a Low-Cost Flash LiDAR}

\author{Peizhao Qiu, Emmanuel A Olowe, Zhenya Zang, Istvan Gyongy, and Danial Chitnis\authormark{*}}

\address{School of Engineering, The University of Edinburgh, Scotland, UK
}

\email{\authormark{*}d.chitnis@ed.ac.uk} 


\begin{abstract*} 

Non-line-of-sight (NLOS) imaging aims to sense objects outside the direct line of sight. While most previous work focuses on the canonical around-the-corner configuration, this work considers a more challenging occlusion-based NLOS setting in which targets are hidden by scene geometry. To avoid the challenge of collecting large real-world datasets, we have developed a photon-based Monte Carlo simulator to generate diverse synthetic training data and validate against experimental data. We trained an AI model to identify multi-bounce photon patterns associated with hidden targets. The model achieves an Average Precision (AP) of 0.91 on synthetic test data and, when transferred directly to 180 real measurements acquired with a TMF8829 SPAD flash LiDAR, achieves 86.67\% accuracy and an AP of 0.95. These results show that simulated transient data can support hidden-target detection with a low-cost commercial sensor and provide an initial step toward more general occlusion-based NLOS sensing and, ultimately, hidden-scene reconstruction.

\end{abstract*}

\section{Introduction}

Non-line-of-sight (NLOS) imaging~\cite{velten2012recovering} aims to detect and localize objects hidden from direct view, with applications in autonomous driving, robotic perception, and search and rescue. Existing approaches include thermal imaging~\cite{maeda2019thermal}, acoustic sensing~\cite{boger2022acoustic}, and ultrafast transient imaging~\cite{xu2021light}. Among these, active time-resolved time-of-flight (ToF) systems offer reduced environmental dependency and high-quality reconstruction performance~\cite{faccio2020non}.
Single-Photon Avalanche Diode (SPAD) LiDAR has emerged as a promising platform for NLOS imaging due to its high temporal resolution and single-photon sensitivity~\cite{faccio2020non,Bronzi2014SPADArray}. By recording time-resolved photon arrivals, a SPAD-based LiDAR captures transient light transport that contains information about hidden scene geometry via indirect reflections. 

Early SPAD-based NLOS reconstruction methods relied primarily on explicit light-transport models that describe multi-bounce propagation between visible relay surfaces and hidden scenes~\cite{velten2012recovering}. Reconstruction is then formulated as an inverse problem and solved using back-projection or optimization-based methods. Subsequent approaches improved reconstruction efficiency and accuracy through techniques such as ellipsoidal back-projection~\cite{Musarra2019NLOS}, FK migration~\cite{lindell2019wave}, and wave-based formulations, including phasor-field methods~\cite{otoole2018confocal}.

Despite these advances, forward model-based methods often rely on simplifying assumptions, such as a single diffuse relay surface and a limited number of light bounces. As more diffuse surfaces are present, light transport becomes increasingly difficult to model analytically, motivating the use of data-driven approaches. More recently, deep learning methods have been introduced to infer hidden scene information directly from transient measurements. Approaches such as DeepNLOS~\cite{grau2020deepnlos} exploit spatio-temporal correlations in transient histograms to implicitly learn multi-bounce light transport without requiring an explicit forward model. While these methods can improve robustness to noise and model mismatch, they typically require large training datasets and may generalize poorly beyond the training distribution.

\begin{figure}[htbp]
\centering
\begin{subfigure}[b]{0.36\linewidth}
  \centering
  \includegraphics[width=\linewidth]{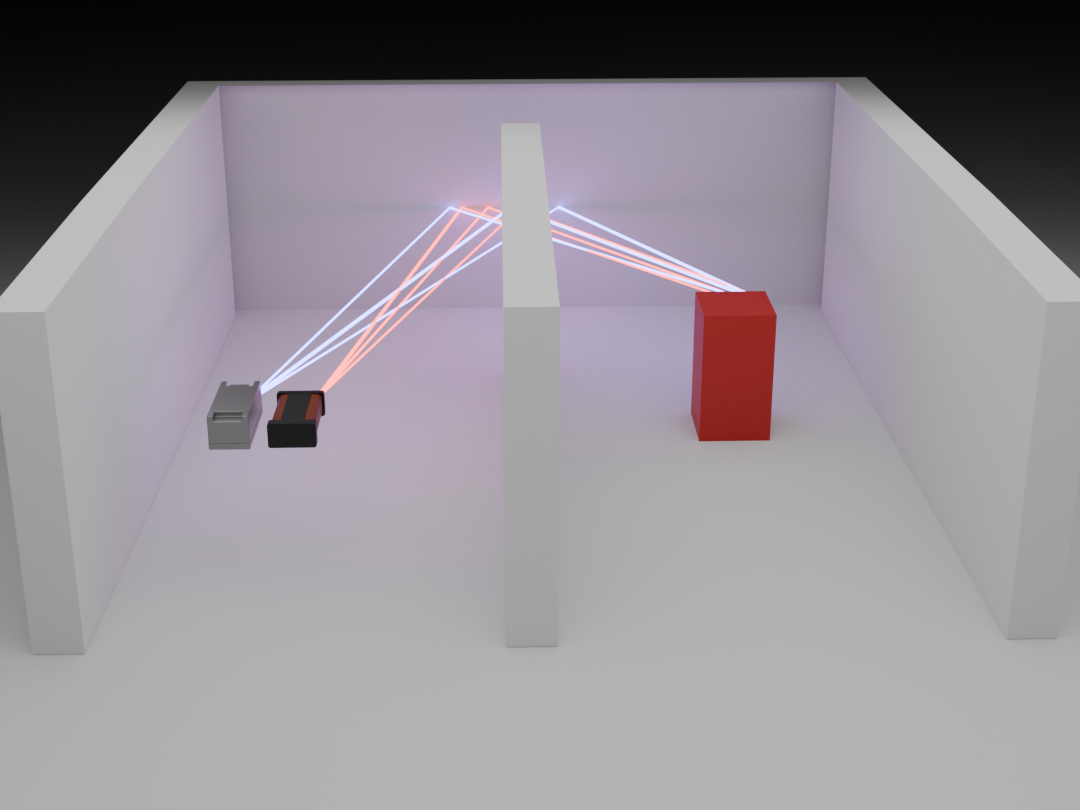}
  \caption{Canonical NLOS case.}
  \label{fig:nlos_comparison_1a}
\end{subfigure}
\hspace{3.5em}
\begin{subfigure}[b]{0.36\linewidth}
  \centering
  \includegraphics[width=\linewidth]{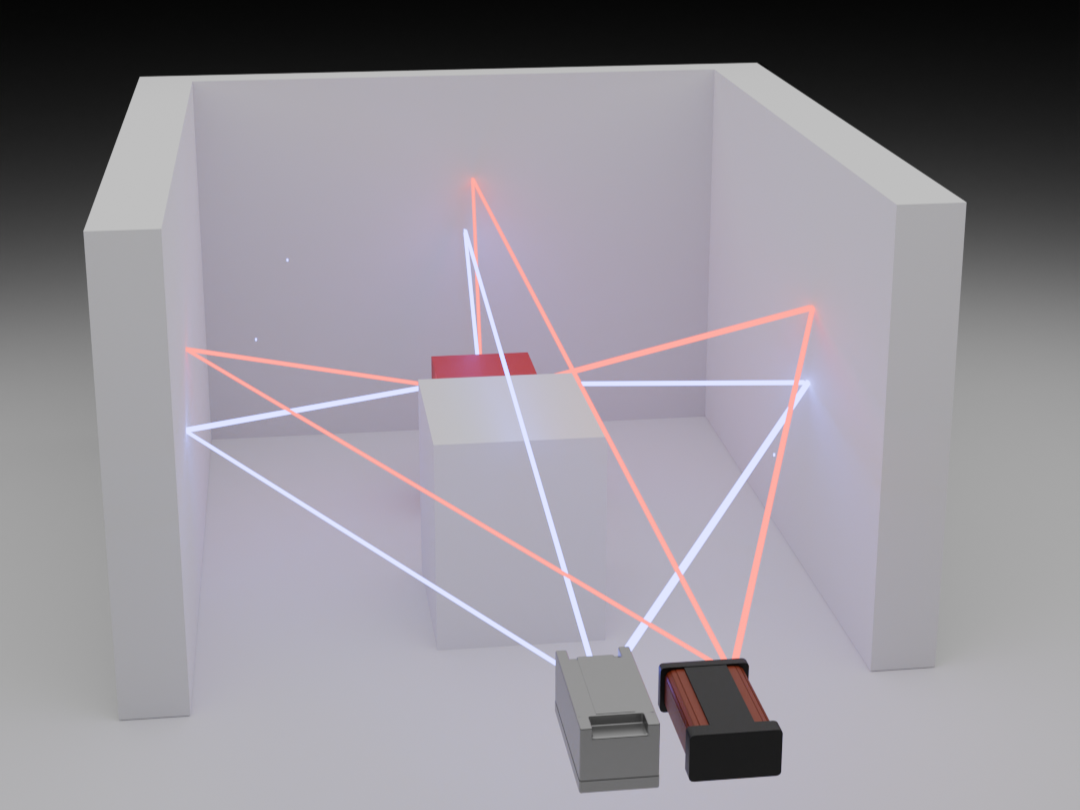}
  \caption{Occlusion-based NLOS case.}
  \label{fig:nlos_comparison_1b}
\end{subfigure}
\caption{(a) Canonical around-the-corner NLOS using a single relay surface, 
and (b) occlusion-based NLOS with complex geometry and unconstrained multi-bounce
paths.}
\label{fig:nlos_comparison_1}
\end{figure}

Most existing NLOS research focuses on the canonical around-the-corner scenario, where light transport is commonly approximated by a three-bounce model~\cite{otoole2018confocal,lindell2019wave}. As illustrated in Fig.~\ref{fig:nlos_comparison_1a}, this configuration typically consists of a single visible diffuse relay surface connecting the detector to a target hidden behind a wall or corner. The known relay geometry constrains the dominant photon propagation paths, resulting in a relatively structured light-transport problem that can be described using established forward models and reconstruction methods. To account for more complex light transport, recent work has extended NLOS imaging beyond the conventional three-bounce assumption~\cite{royo2023virtual}, although these approaches largely remain within relay-based scenarios. In contrast, a more ill-posed, occlusion-based NLOS scenario is considered, illustrated in Fig.~\ref{fig:nlos_comparison_1b}, in which a target is hidden behind an occluder in a semi-enclosed environment. Unlike conventional NLOS setups that rely on a single relay wall, photons can interact with multiple surrounding surfaces, producing dense, overlapping paths with different reflection orders, path lengths, and attenuation levels. As a result, the weak target-dependent signal is difficult to isolate from the transient response. This work makes three main contributions: we study a more challenging occlusion-based NLOS setting, develop an accelerated Monte Carlo framework for photon transport and SPAD noise simulation, and train an ML-based classifier entirely on synthetic data, then validate it on a low-cost commercial SPAD LiDAR.

\section{Methodology}\label{sec:methodology}
SPAD-based LiDAR measures photon time-of-flight (ToF) by emitting light into a scene and recording the time delay of the return signal. It can operate in flash or scanning modes: flash LiDAR uses wide-field illumination and a 2D SPAD array for fast scene capture, while scanning LiDAR sweeps a focused beam to achieve higher angular resolution~\cite{otoole2018confocal}. In both cases, photon arrivals are accumulated into a 3D spatio-temporal histogram~\cite{shin2016photonefficient}, where the temporal bin width determines depth precision~\cite{McCarthy2013LongRange}. Using avalanche breakdown and a time-to-digital converter (TDC), SPADs provide high sensitivity and photon efficiency~\cite{Bronzi2014SPADArray}. 
In this work, a low-cost flash LiDAR is used to experimentally validate the proposed method. Compared with task-specific scanning LiDAR systems, flash LiDAR distributes its emitted optical power over a wide field of view, resulting in lower per-pixel photon counts and potentially greater sensitivity to sensor noise and spatial non-uniformity~\cite{villa2021spads}.

\begin{figure}[htbp]
\centering
  \includegraphics[width=0.6\linewidth]{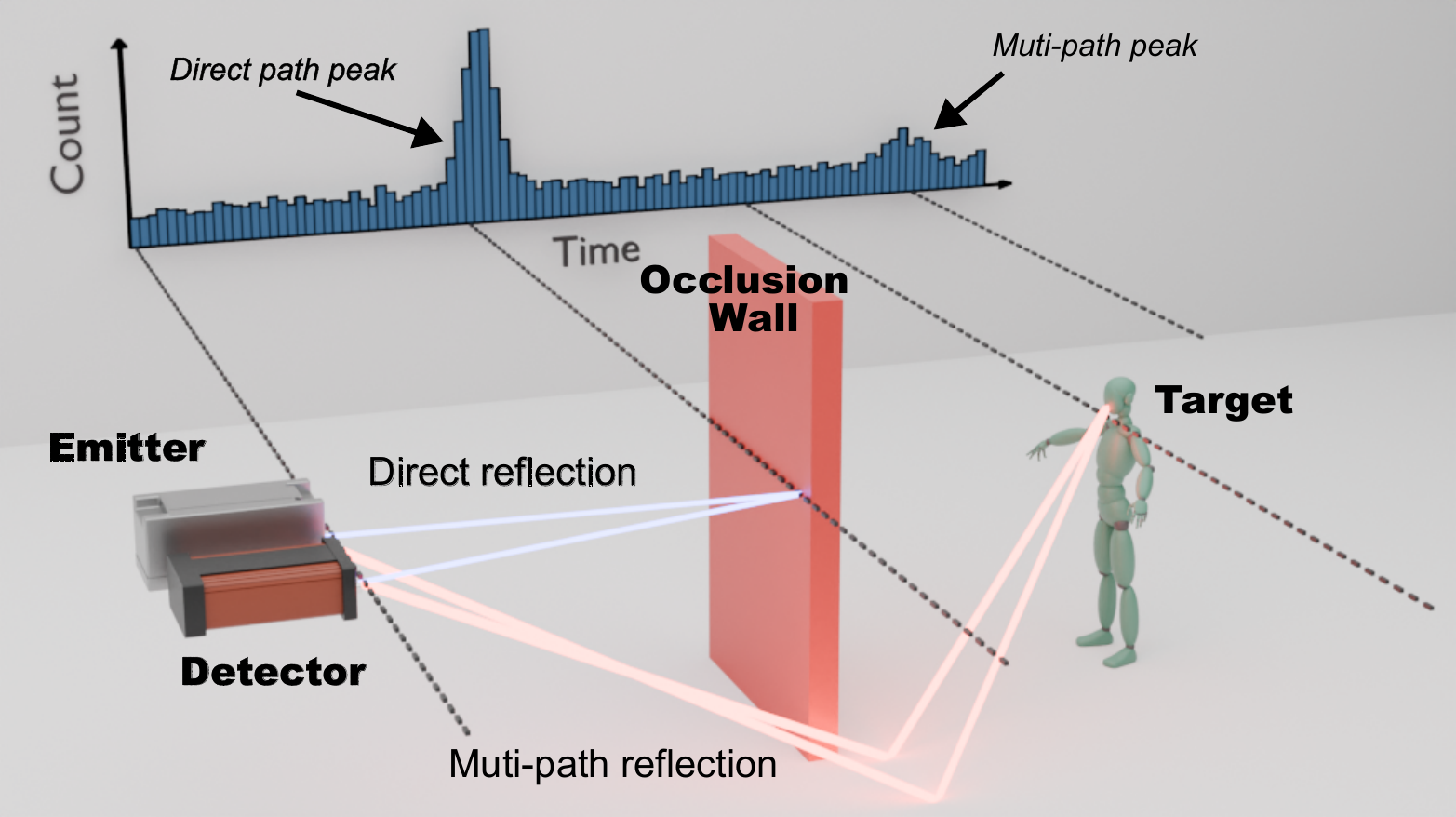}
\caption{ In a SPAD transient histogram, the direct-path peak indicates the presence of occlusion, while the embedded multipath peaks encode NLOS information.}
\label{fig:NLOS_histogram}
\end{figure}

During photon propagation, most detected photons undergo direct reflections, producing strong early peaks in the transient image, while a smaller fraction undergo multiple reflections, interact with hidden surfaces, and return as weaker, delayed responses. These multi-bounce photons may carry NLOS information and often appear as subtle, delayed ripples in the transient histogram, as shown in Fig.~\ref{fig:NLOS_histogram}. However, their detection probability is exponentially lower due to additional scattering and absorption losses. Therefore, it may be possible to extract NLOS information from SPAD transient histograms. Such information can be accurately modeled using Monte Carlo simulation, which samples photon paths while accounting for visibility, multiple reflections, surface interactions, and propagation attenuation. This preserves weak NLOS signals from complex multi-bounce paths, which are difficult to model with analytical equations in scenes with arbitrary occlusions and high-order reflections.

\begin{figure}[htbp]
\centering
  \includegraphics[width=0.8\linewidth]{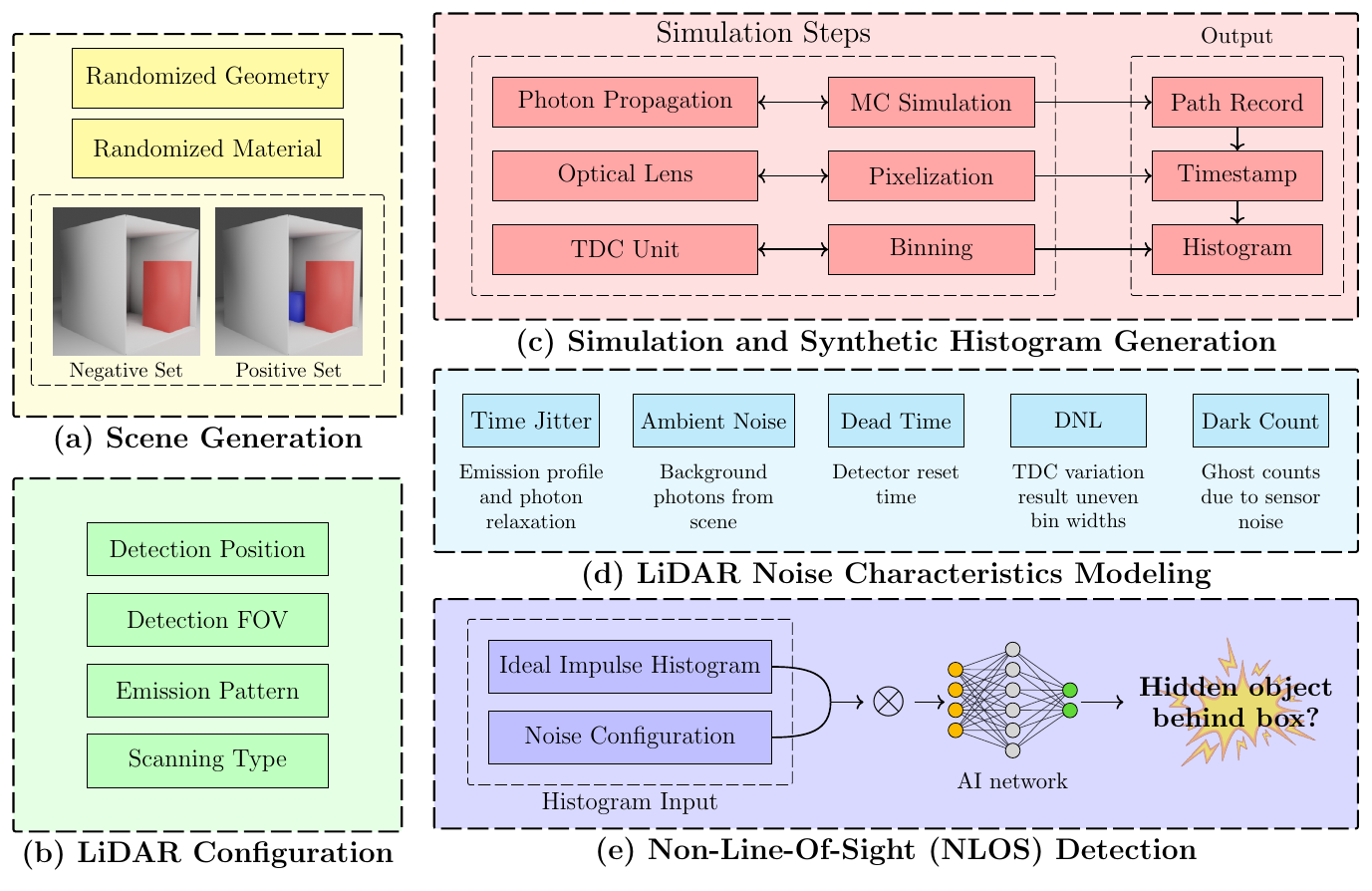}
\caption{Pipeline for synthetic data generation and NLOS detection framework.}
\label{fig:system_flowchart}
\end{figure}

As discussed in the previous section, occlusion-based NLOS sensing is more challenging than the conventional around-the-corner scenario because explicit light-transport modeling becomes difficult in complex scenes. Data-driven methods provide an alternative by learning NLOS features directly from transient measurements. However, the performance of data-driven methods strongly depends on the quality and diversity of the training data. No existing dataset is available for this setting, and collecting sufficient experimental data is impractical due to budget and resource constraints. Therefore, the model is trained entirely on synthetic data generated using the simulation-driven pipeline illustrated in Fig.~\ref{fig:system_flowchart}. The pipeline generates positive samples with a hidden object present and negative samples without one. Each scene is generated with randomized geometry and material properties to improve generalization. The virtual SPAD system is parameterized by sensor position, emission pattern, scanning mode, and field of view. Photon transport is simulated using a Monte Carlo framework, followed by pixelization and temporal binning to form ideal transient histograms. Noise effects are then incorporated to produce realistic device-level SPAD histograms. The resulting transient histograms are used to train a custom neural network for binary NLOS detection. The model learns target-dependent features from the simulated multi-bounce photon responses and predicts whether a hidden object is present or absent.

\section{SPAD simulation}

The SPAD simulation has two stages: Monte Carlo photon propagation and sensor noise modeling. Photon transport is simulated from scene geometry and surface properties to generate arrival times, which are
converted into transient histograms. The noise model then incorporates ambient noise, the instrument
response function (IRF), detector dead time, TDC differential non-linearity (DNL), and sensor-specific fixed-pattern noise to better match real SPAD measurements~\cite{hernandez_computational_2017,greeley2019characterizing,villa2021spads}.

\subsection{Monte Carlo simulation and acceleration}

To capture multi-path photon transport carrying NLOS information, a Monte Carlo simulation was used to model photon propagation. Photon absorption, reflection, and propagation direction were treated as stochastic processes determined by scene geometry and surface properties. SYCL~\cite{sycl2020} provides heterogeneous hardware acceleration, while a Bounding Volume Hierarchy (BVH)~\cite{pharr2023pbrt} accelerates ray--surface intersection tests. Given a virtual scene and parameterized SPAD configuration, the simulator generates photon arrival timestamps and maps them to image pixels according to the LiDAR optical model, producing ideal transient histograms.

\subsection{SPAD noise simulation }
Following the simulation of photon arrival timestamps from scene propagation, a SPAD noise model is applied to transform the ideal photon measurements into realistic observations that account for device-level characteristics. Four noise sources are considered: ambient noise, SPAD IRF, dead-time, and DNL distortion from the TDC unit.

\subsubsection{Ambient Noise Simulation}

\begin{figure}[htbp]
\centering

\makebox[\linewidth][c]{%
\subcaptionbox{Synthetic scene image.
\label{fig:spad_simulation_rendering}}{%
  \includegraphics[height=3cm]{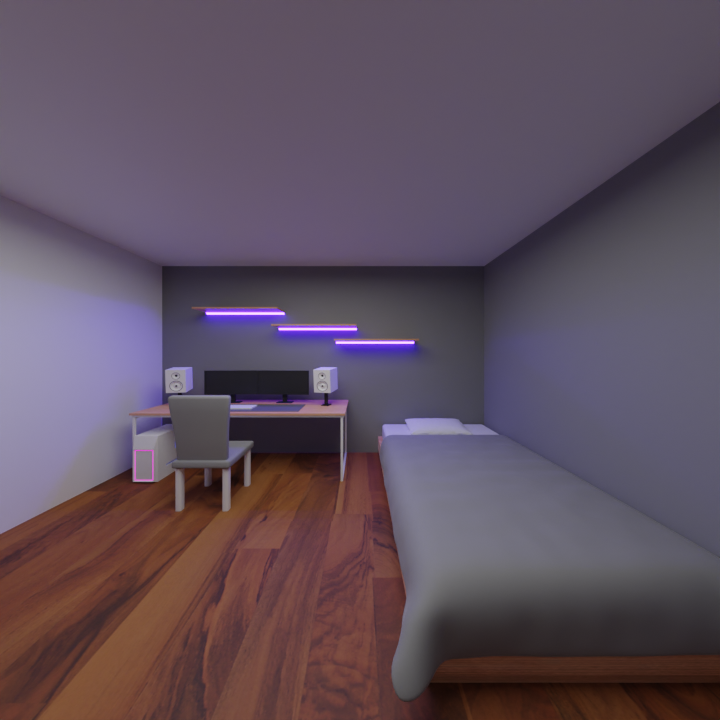}%
}
\hspace{1.3cm}
\subcaptionbox{Scene reflectivity map.
\label{fig:spad_simulation_reflective}}{%
  \includegraphics[height=3cm]{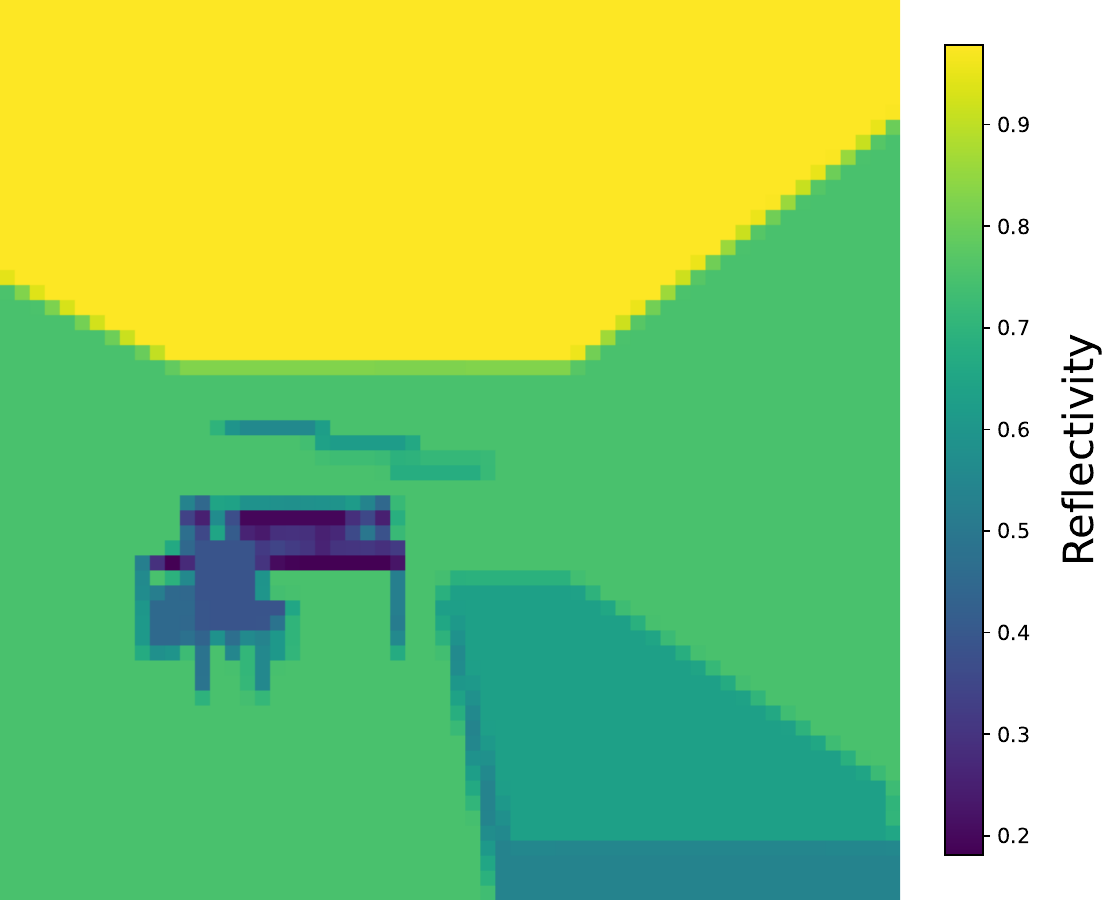}%
}
\hspace{0.7cm}
\subcaptionbox{Scene depth image.
\label{fig:spad_simulation_depth}}{%
  \includegraphics[height=3cm]{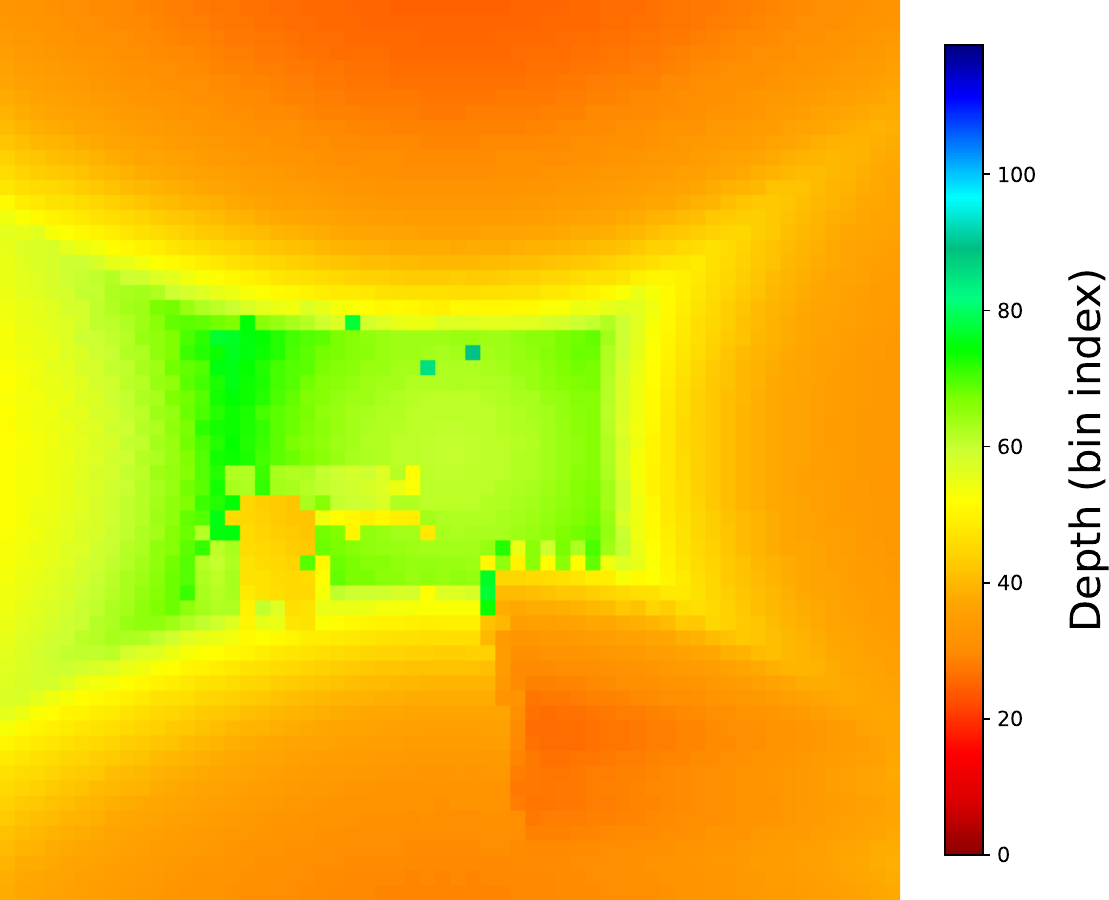}%
}
}

\caption{Rendered scene~(a) with corresponding reflectivity map~(b) and depth image~(c).}
\label{fig:ambient_simulation}
\end{figure}

Ambient noise originates from background illumination and is independent of the emitted LiDAR pulse. For each pixel $(x,y)$, ambient photon arrivals are modeled as a Poisson process:
\begin{equation}
N_{\text{ambient}}(x,y,t)
\sim
\mathrm{Poisson}\!\left(\lambda_a(x,y)\right),
\qquad
\lambda_a(x,y)=k\,\rho(x,y).
\end{equation}
where $\lambda_a(x,y)$ is the mean ambient photon count, approximated under 
isotropic illumination from the first-hit reflectivity map. Here, $\rho(x,y)$ 
denotes the surface reflectivity, and $k$ controls the overall ambient level. For the scene shown in Fig.~\ref{fig:ambient_simulation}(a), the pixel-wise ambient photon level is determined from the reflectivity map in Fig.~\ref{fig:ambient_simulation}(b). Since ambient photons are uncorrelated with the laser pulse, their arrival times are modelled as uniformly distributed over the temporal window. The resulting transient histogram, formed from the simulated LiDAR return and ambient photons, is then used to reconstruct the depth image as shown in Fig.~\ref{fig:ambient_simulation}(c) using centroid-based peak estimation, which estimates the photon time-of-flight from the weighted centre of the histogram peak~\cite{villa2021spot}

\subsubsection {SPAD system IRF modeling}
In this work, the SPAD system IRF is modeled using an exponentially modified Gaussian (EMG)~\cite{greeley2019characterizing}. The EMG combines a Gaussian component describing the symmetric temporal spread of the LiDAR response with an exponential component representing its asymmetric decay, thereby capturing timing jitter and the electronic relaxation response of the SPAD system. Their convolution provides a compact model of the asymmetric temporal response observed in photon-counting LiDAR measurements~\cite{hernandez_computational_2017}.

\begin{equation}
t_j \sim \mathcal{N}(\mu,\sigma^2) \ast \mathrm{Exp}(\tau)
\end{equation}

The timing delay $t_j$ is sampled from this distribution and added to the simulated photon arrival time. where $\mu$ denotes the systematic timing offset, $\sigma$ the standard deviation of the Gaussian component, and $\tau$ the exponential relaxation constant.

\subsubsection{Dead-time}
\begin{figure}[htbp]
\centering
\begin{subfigure}[b]{0.48\linewidth}
  \centering
  \includegraphics[width=\linewidth]{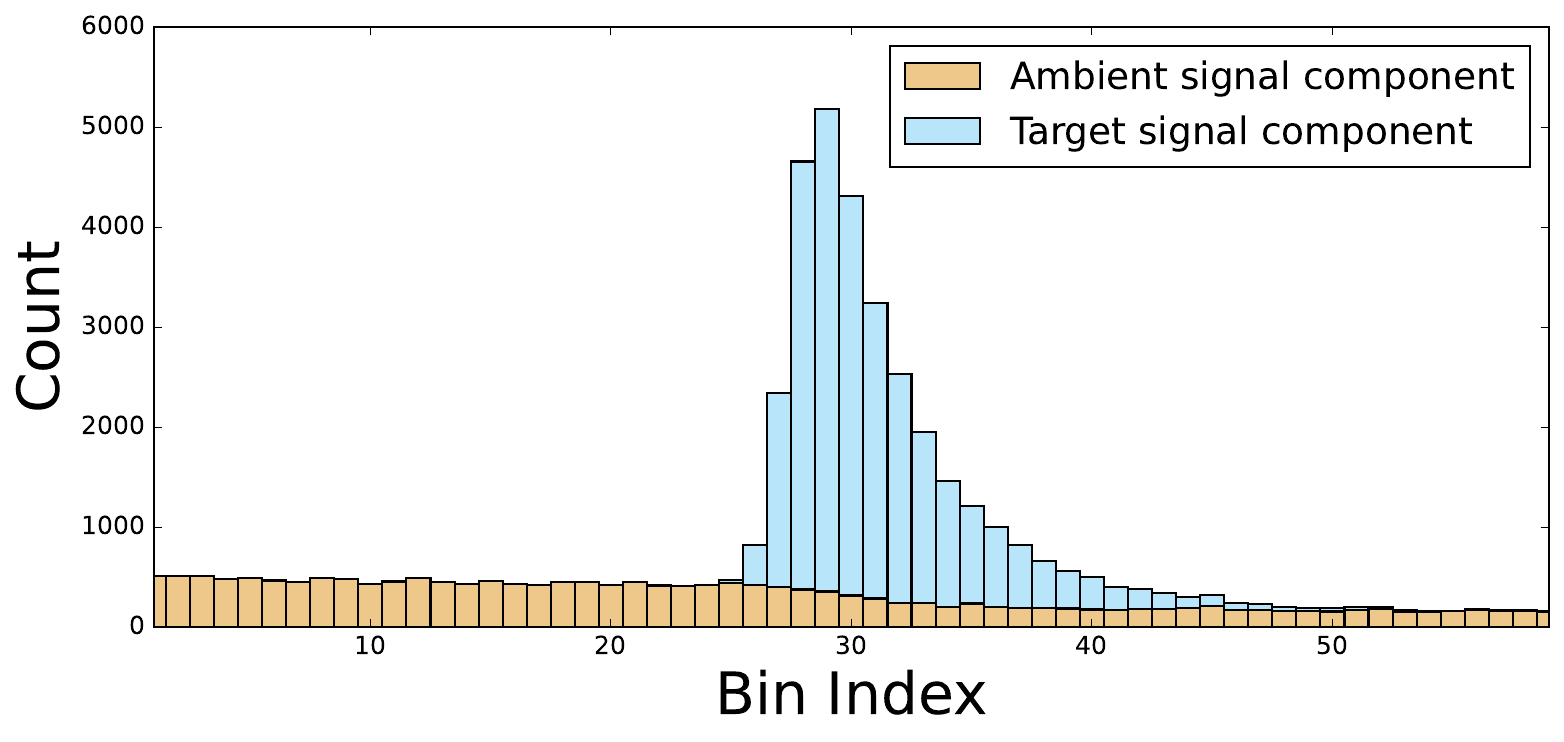}
  \caption{Low ambient flux histogram.}
  \label{fig:low_flux}
\end{subfigure}
\hfill
\begin{subfigure}[b]{0.48\linewidth}
  \centering
  \includegraphics[width=\linewidth]{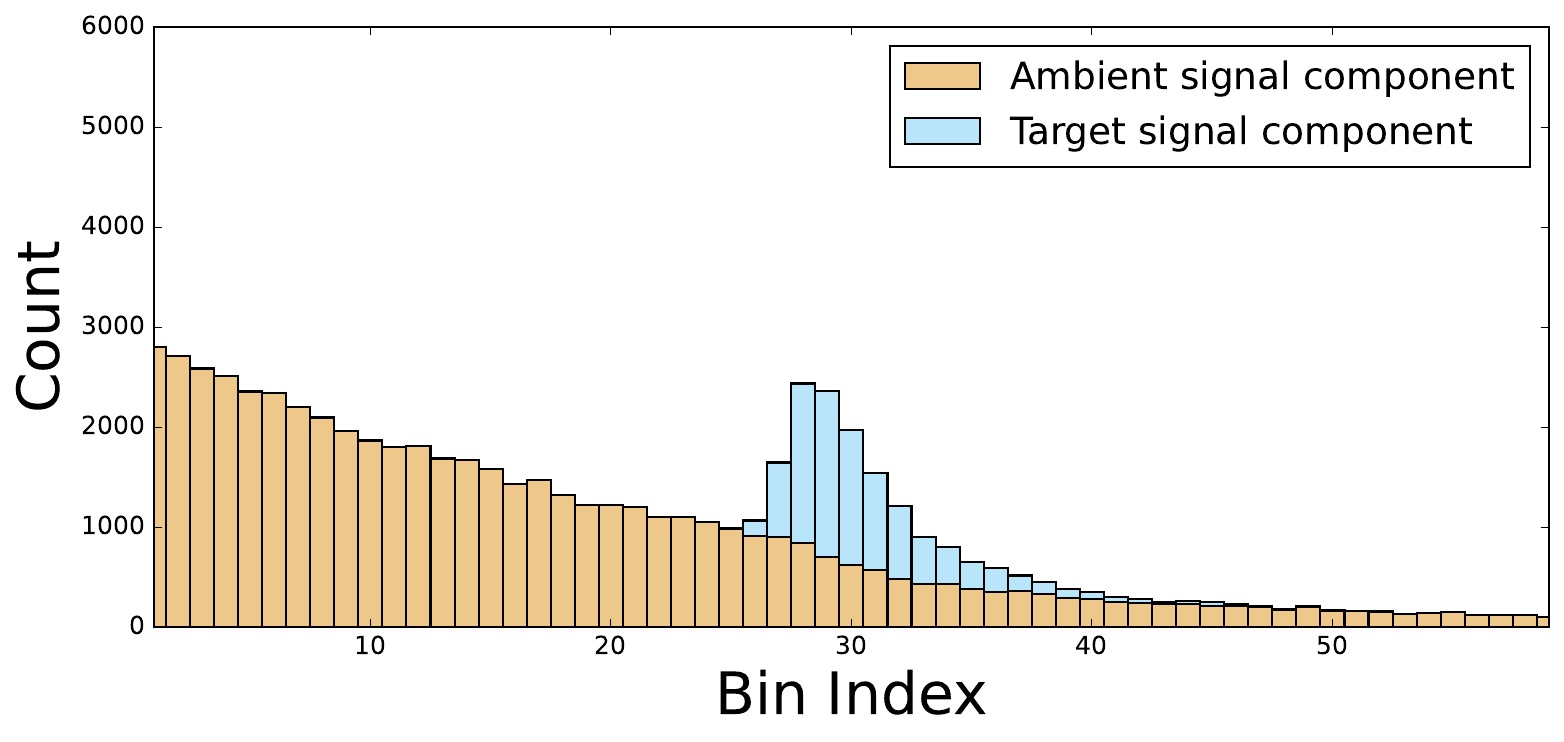}
  \caption{High ambient flux histogram.}
  \label{fig:high_flux}
\end{subfigure}
\caption{
Minor pile-up in low ambient flux~(a) and severe
pile-up in high ambient flux~(b).
}
\label{fig:pile_up}
\end{figure}

Dead-time is the short recovery period after a SPAD detects a photon, during which it cannot register another~\cite{Rapp2021HighFlux}. This can cause pile-up, where early photon detections suppress later arrivals and bias the transient histogram toward shorter arrival times~\cite{rahmanpour2024implementable}. In this work, a Monte Carlo algorithm is used to model SPAD dead-time. As shown in Fig.~\ref{fig:pile_up}, under low ambient flux [Fig.~\ref{fig:pile_up}(a)], the histogram maintains a nearly uniform background and preserves the target signal. Under high ambient flux [Fig.~\ref{fig:pile_up}(b)], first-photon detection and dead-time shift the ambient background toward earlier bins of the TDC, producing a decreasing temporal slope and suppressing later main-reflection and multi-path signals.

\subsubsection{DNL}

DNL noise is introduced during the binning stage of the TDC, where photon timestamps are converted into a time-resolved histogram. This effect arises from differential nonlinearity in the TDC, where the actual bin widths deviate from the ideal uniform spacing, leading to non-uniform temporal quantization and artificial bin-to-bin variations in the measured histogram.  In this simulation, DNL is modeled using a synthetic 3D matrix representing the relative variation in histogram bin widths. Values above or below one indicate wider or narrower bins, respectively, with the degree of variation controlled by a user-defined standard deviation.

\subsection{Impact of noise on NLOS information}

\begin{figure}[htbp]
\centering
  \includegraphics[width=\linewidth]{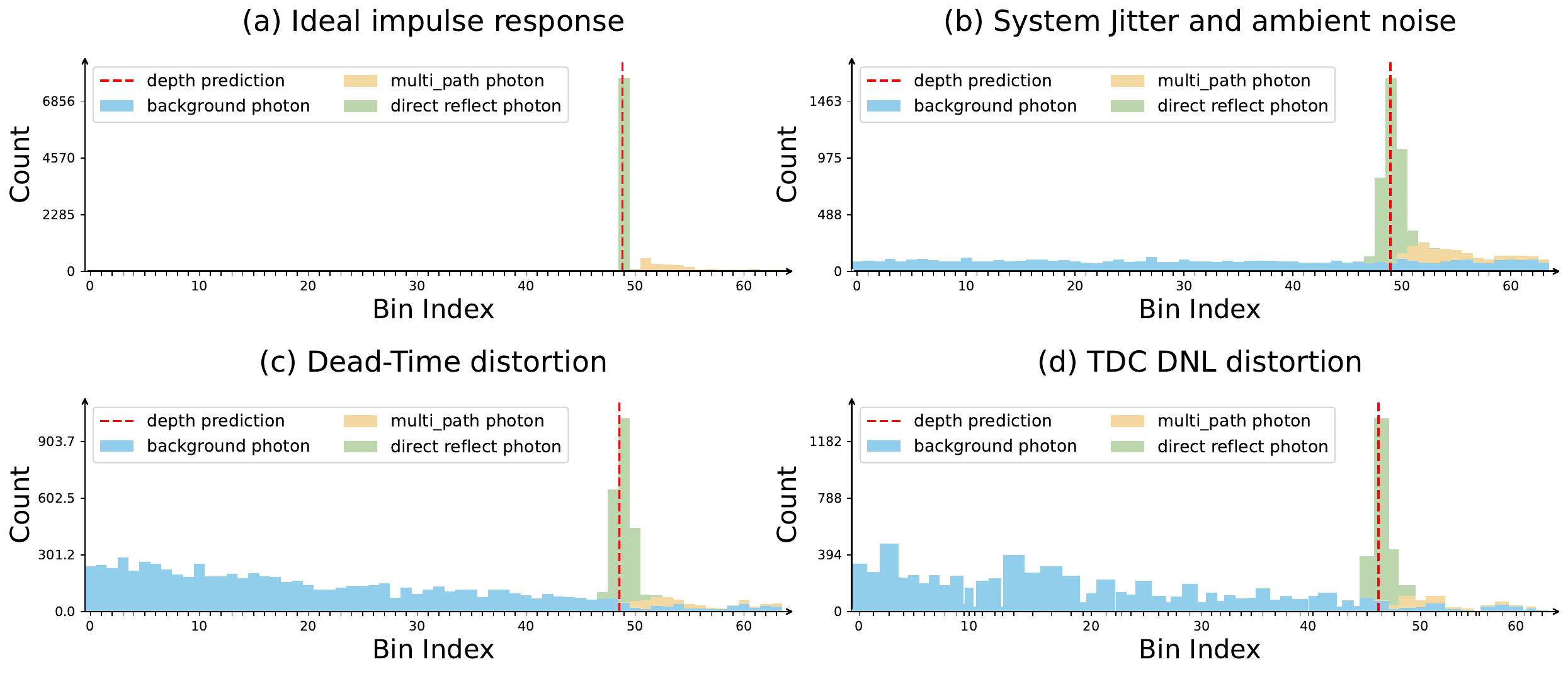}
\caption{Influence of different noise distortions on the transient histogram.}
\label{fig:overall_histogram_plot}
\end{figure}

As discussed in Section~\ref{sec:methodology}, NLOS information is encoded in multi-path photon arrivals. Fig~\ref{fig:overall_histogram_plot} illustrates how different noise sources affect this information in the transient histogram. In the ideal case, as shown in Fig~\ref{fig:overall_histogram_plot}~(a), the direct reflection forms a sharp peak, clearly separated from the weaker multipath signal. With time jitter and ambient noise Fig~\ref{fig:overall_histogram_plot}~(b), the main peak broadens and the background level increases, although the multipath tail remains distinguishable. Under high photon flux, Fig~\ref{fig:overall_histogram_plot}~(c), dead time suppresses later photon arrivals following early detections, reducing the signal-to-noise ratio for both the direct and multi-path components. Finally, DNL Fig~\ref{fig:overall_histogram_plot}~(d) introduces bin-to-bin variations that further mix the multipath signal with the background and main reflection, making the NLOS component more difficult to distinguish.

\subsection{Sensor calibration and pixel-pattern noise}
\begin{figure}[htbp]
\centering
  \includegraphics[width=\linewidth]{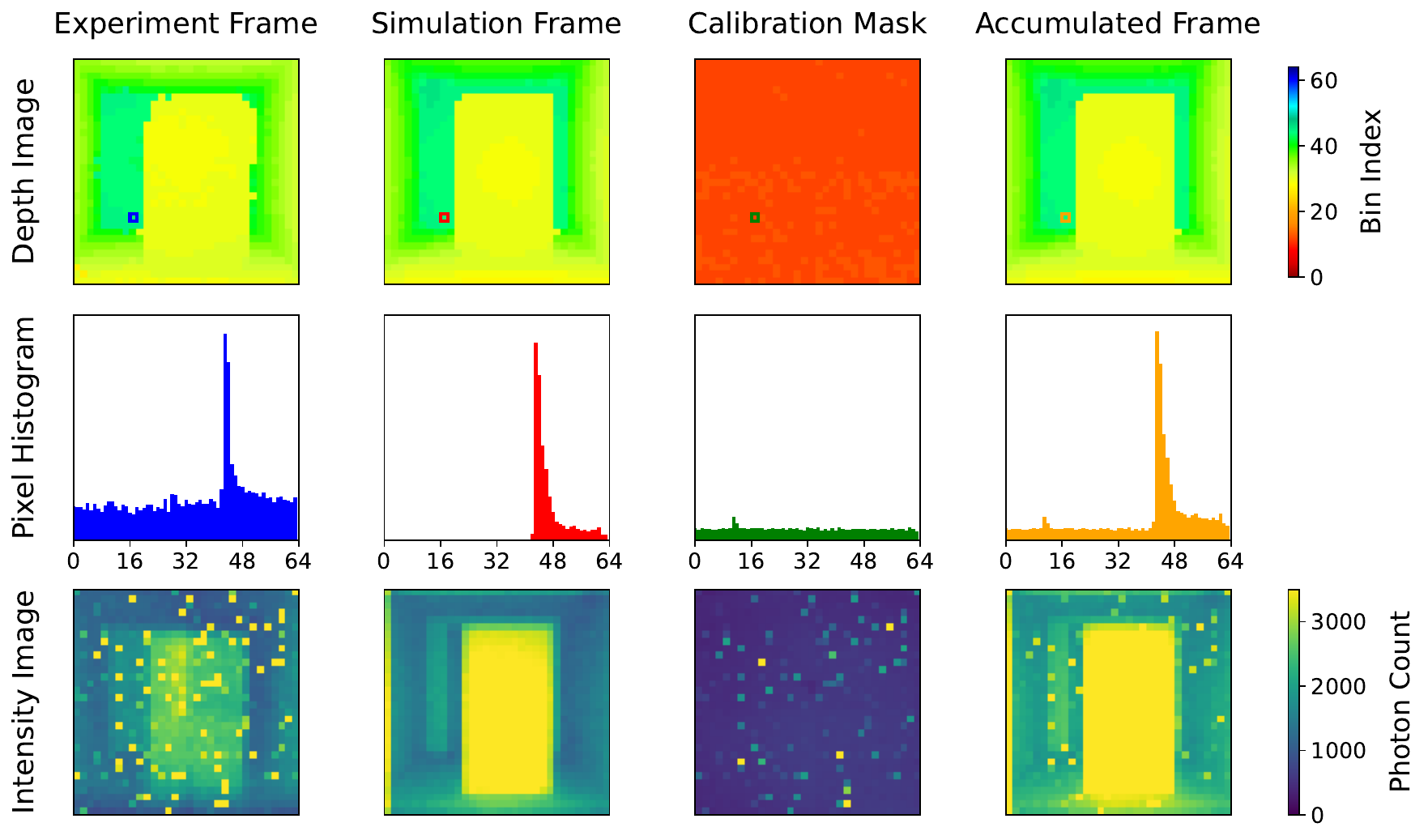}
\caption{Comparison of experimental, simulated, calibration, and accumulated results. Each column shows the depth image, pixel histogram, and intensity image. }
\label{fig:calibration_process}
\end{figure}

The simulation reproduces the scene impulse response while modeling the main LiDAR noise sources. Some effects are difficult to describe with a general analytical model, particularly in flash LiDAR, where multi-pixel detector arrays and shared optical and readout paths introduce spatially varying responses. As a result, dark counts, cross-talk, and after-pulsing can differ across pixels and sensor units, making a single set of noise parameters insufficient.
To capture this sensor-specific fixed-pattern noise, a calibration mask was obtained by covering the LiDAR emitter and recording the sensor response. The measured noise map is then added to the simulated transient histograms. Although it does not reproduce each pixel histogram exactly, it better matches the spatial photon-count distribution of the real sensor, as shown in Fig.~\ref{fig:calibration_process}. This calibration helps reduce the simulation-to-experiment gap~\cite{lee2023caspi}. Models trained without it failed to generalize, while masks with similar global statistics but different spatial patterns also performed poorly, indicating that matching the spatial noise distribution is important for simulation-to-experimental transfer.

\section{Detection model architecture}

\begin{figure}[htbp]
\centering
  \includegraphics[width=\linewidth]{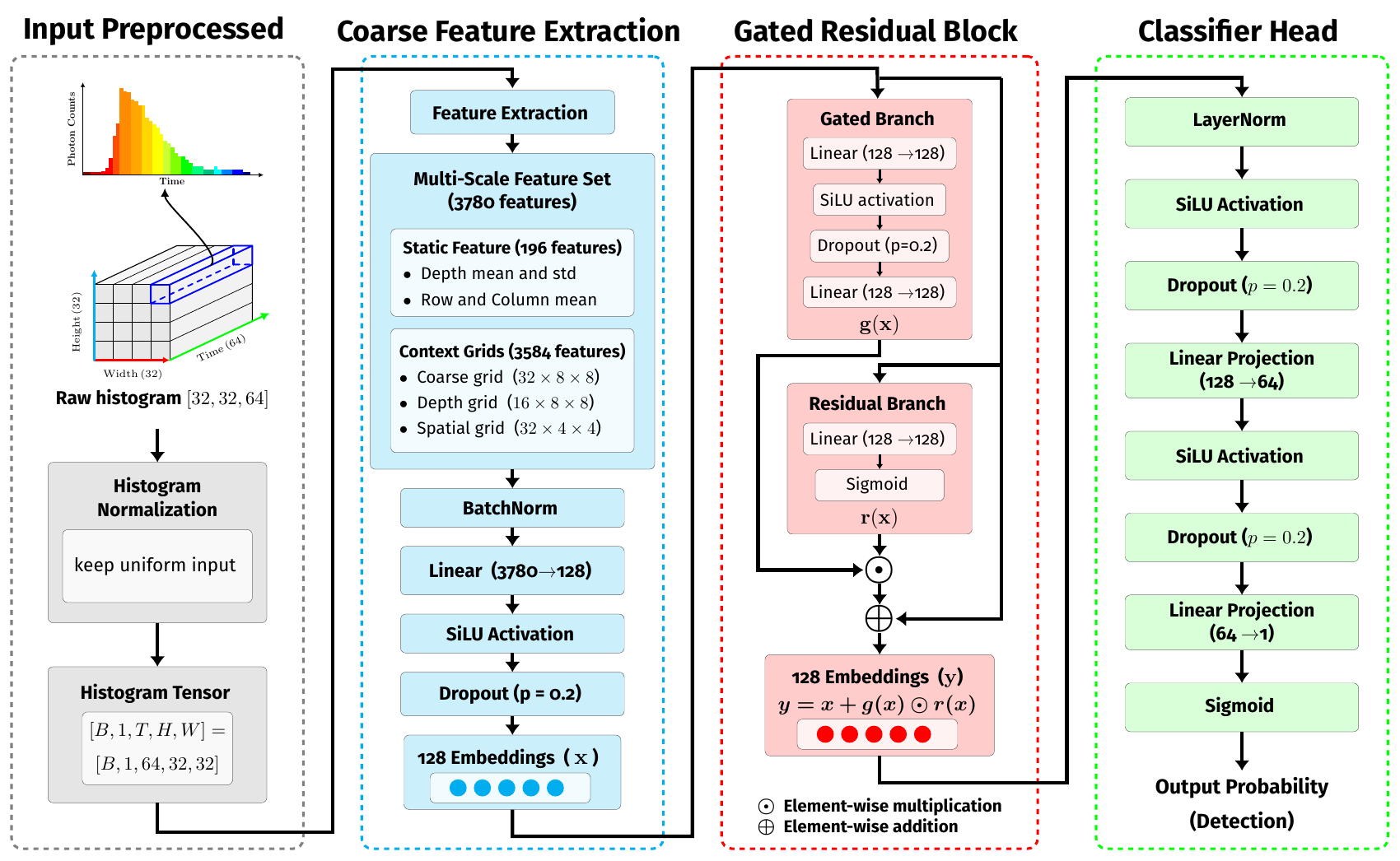}
\caption{Workflow of the CSDP-MLP architecture. }
\label{fig:deep_learning_architecture}
\end{figure}

While most machine-learning-based NLOS methods focus on hidden-scene reconstruction, this work targets the detection of fully occluded objects. A custom Coarse Spatial-Depth Projection MLP (CSDP-MLP) was therefore developed that takes LiDAR transient histograms with a spatial dimension of $(32\times32)$ and a temporal dimension of 64, i.e., ([32,32,64]), as input and outputs the probability that a hidden object is present, as shown in Fig~\ref{fig:deep_learning_architecture}. Before being fed into the network, each raw histogram $\boldsymbol{H}\in\mathbb{R}^{32\times32\times64}$ is normalized into a numerically consistent form for training~\cite{Gutierrez-Barragan_2023_ICCV,Hong2023ImageFree}. The complete normalization process is summarized as:

\begin{equation}
\boldsymbol{H}_{\mathrm{scale}}
=
\boldsymbol{H}
\frac{C}{\sum_{r,c,b} H_{r,c,b}},
\qquad
\boldsymbol{H}_{\log}
=
\log\!\left(1+\boldsymbol{H}_{\mathrm{scale}}\right),
\qquad
\boldsymbol{H}_{\mathrm{norm}}
=
\frac{\boldsymbol{H}_{\log}-\mu_{\mathrm{train}}}
{\sigma_{\mathrm{train}}}.
\end{equation}

The input transient histogram $\boldsymbol{H}$ is first rescaled to a fixed total photon count $C$, producing the photon-count-normalized histogram $\boldsymbol{H}_{\mathrm{scale}}$ to reduce photon-flux variation across samples. A logarithmic transformation is then applied to obtain $\boldsymbol{H}_{\log}$, compressing the photon-count dynamic range while preserving weaker signals. The resulting histogram is standardized using the training-set mean $\mu_{\mathrm{train}}$ and standard deviation $\sigma_{\mathrm{train}}$, yielding the final normalized histogram $\boldsymbol{H}_{\mathrm{norm}}$ to improve numerical consistency and training stability. Finally, the normalized histograms are grouped into mini-batches with tensor layout $[B,1,64,32,32]$ and fed into the network.

A coarse feature extraction stage then converts the high-dimensional, sparse histogram tensor into more discriminative, compact feature embeddings. This reduces dimensionality and computational cost while improving training stability and limiting overfitting~\cite{jia2022feature}. The histogram is first projected into multi-scale features composed of global statistics and context grids. Global statistics summarize the overall photon distribution, while context grids aggregate the histogram at different spatial and temporal resolutions to retain both local and broader scene structure. The multi-scale features are then processed by batch normalization~\cite{huang2023normalization}, a fully connected layer, and SiLU activation to form a one-dimensional embedding. Batch normalization stabilizes the feature distribution, while SiLU activation~\cite{elfwing2018silu} introduces nonlinearity. The resulting compact embedding is then used for subsequent processing.

The gated residual MLP block~\cite{srivastava2015highway} is introduced to further refine the compact embedding. It preserves useful information from the original features through the residual path, while a parallel gate controls how much transformed information is added~\cite{he2016deep}. This selective update allows the network to improve the feature representation without discarding previously learned information, supporting more stable optimization and better generalization. Finally, the classifier head block maps the feature embedding to an NLOS detection probability using two fully connected layers with SiLU activation and layer normalization~\cite{huang2023normalization} and the final classification score is converted into an NLOS detection probability with a sigmoid function. 

\section{Evaluation Metrics}
The NLOS detection probability output from the network is then thresholded to obtain a binary prediction of whether the hidden object is present. The performance of the network is evaluated using accuracy, precision, recall, F1-score, AUC, and BCE loss. Accuracy measures the overall proportion of correct predictions. Precision measures the ratio of correctly detected hidden-object samples to all samples predicted as positive, while recall measures the ratio of correctly detected hidden-object samples to all actual positive samples, including missed detections. F1-score is the harmonic mean of precision and recall, balancing false and missed detections. AUC measures class separation across all decision thresholds, with higher values indicating better discrimination. BCE loss measures the difference between predicted probabilities and ground-truth labels, with lower values indicating better probabilistic predictions. The model is further analyzed using the precision--recall curve and confusion matrix. The precision--recall curve shows the trade-off between precision and recall across different decision thresholds, while average precision (AP) summarizes performance over the full curve. The confusion matrix reports the numbers of true positives, true negatives, false positives, and false negatives, providing a direct view of class-specific prediction errors.

\section{Experiment and Results}

The proposed NLOS classification model was trained entirely on synthetic data and validated using real experimental measurements. Performance was evaluated on both synthetic and experimental datasets to assess classification accuracy and generalization from simulation to experiment. This section first presents the experimental setup and data acquisition, then describes training and evaluation on synthetic data, and finally validates on real measurements to assess transfer performance and remaining practical limitations.

\subsection{Experiment setup}

\begin{figure}[htbp]
\centering
  \includegraphics[width=0.9\linewidth]{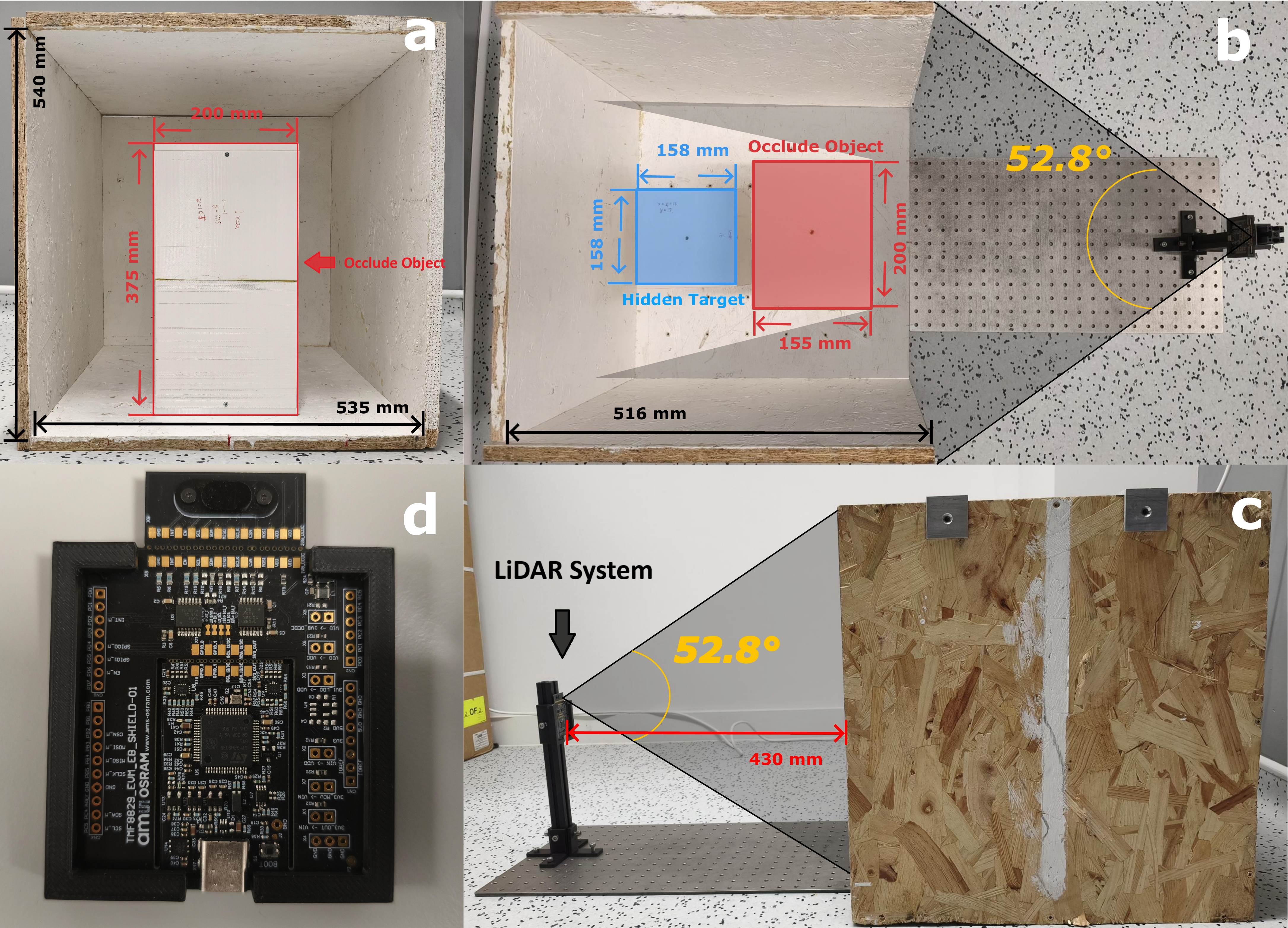}

\caption{Experimental setup from different viewpoints: (a) front view, (b) top view, (c) side view, and (d) TMF8829 dToF LiDAR sensor.}

\label{fig:experiment_set_up}
\end{figure}

The experiment is set up inside a $516\times535\times540$~mm box with white-painted interior surfaces for high diffuse reflectivity. A $155\times200\times375$~mm occluder blocks the direct view of a $158\times158\times168$~mm target positioned behind it. The sensor is fixed on an optical plate and directed toward the scene. The experiment uses a commercially available ams TMF8829 dToF SPAD sensor with a $67.9^\circ\times52.8^\circ$ horizontal and vertical FoV. It supports resolutions from $8\times8$ to $48\times32$ and temporal bin widths of 200--1600~ps. In this experiment, the $48\times32$ high-accuracy short-range mode with 64 bins per pixel and a 200~ps bin width was used. The spatial resolution is cropped to $32\times32$, corresponding to a $52.8^\circ\times52.8^\circ$ FoV, to match the square experimental region of interest. The complete setup is shown in Fig.~\ref{fig:experiment_set_up}.

\subsection{Synthetic data generation and training}

A total of 38,000 synthetic samples were generated over two weeks using an NVIDIA GPU for model training. The training samples were evenly split between samples with a hidden object present and samples with a hidden object absent. To improve robustness and reduce overfitting, the dataset varied scene geometry, material properties, LiDAR position, target and occluder size and position, photon flux, and sensor noise parameters. The LiDAR orientation and FoV were kept fixed to match the experimental setup, with the sensor perpendicular to the scene and the occluder always large enough to fully block the target from direct view. This preserves the experimental geometry while providing sufficient variation to support generalization.

\begin{table}[t]
\centering
\caption{Variation in noise parameters used to generate synthetic training data.}
\begin{tabular}{lclc}
\hline
Parameter & Range & Parameter & Range \\
\hline
Acquisition cycles & $25000\sim32000$
& TDC DNL variance & $5\%\sim10\%$ \\

IRF std~($\sigma$) & $10\sim100$ ps
& Calibration counts ratio & $20\%\sim160\%$ \\

IRF timing offset~($\mu$) & $2300\sim2500$ ps
& Ambient noise ratio & $10\%\sim150\%$ \\

IRF relaxation~($\tau$) & $50\sim350$ ps
& & \\
\hline
\end{tabular}
\label{table:noise_parameters}
\end{table}

Additional diversity was introduced by varying the sensor noise parameters, as shown in Table~\ref{table:noise_parameters}. A total of 36 noise configurations were used, with 1,000 samples generated for each configuration. The parameter ranges were based on the sensor's actual characteristics, with additional variation to improve robustness under different measurement conditions. The training set also included 2,000 clean samples containing only multipath effects. These samples preserve the underlying NLOS signal and help the model separate scene-dependent multipath features from sensor-induced noise. During training, D4 spatial augmentation, consisting of four right-angle rotations and their mirrored counterparts, was applied to the spatial dimensions of each training sample to improve robustness to object orientation and mirror symmetry. For each histogram, one of eight transformations was randomly selected: rotations of $0^\circ$, $90^\circ$, $180^\circ$, and $270^\circ$, with or without reflection. This increases training diversity and reduces dependence on specific spatial orientations.
The proposed NLOS classifier was trained for 50 epochs on the synthetic dataset, showing stable convergence, closely aligned training and validation losses, and no evident overfitting.

\subsection{Model performance}

\begin{figure}[htbp]
\centering
  \includegraphics[width=\linewidth]{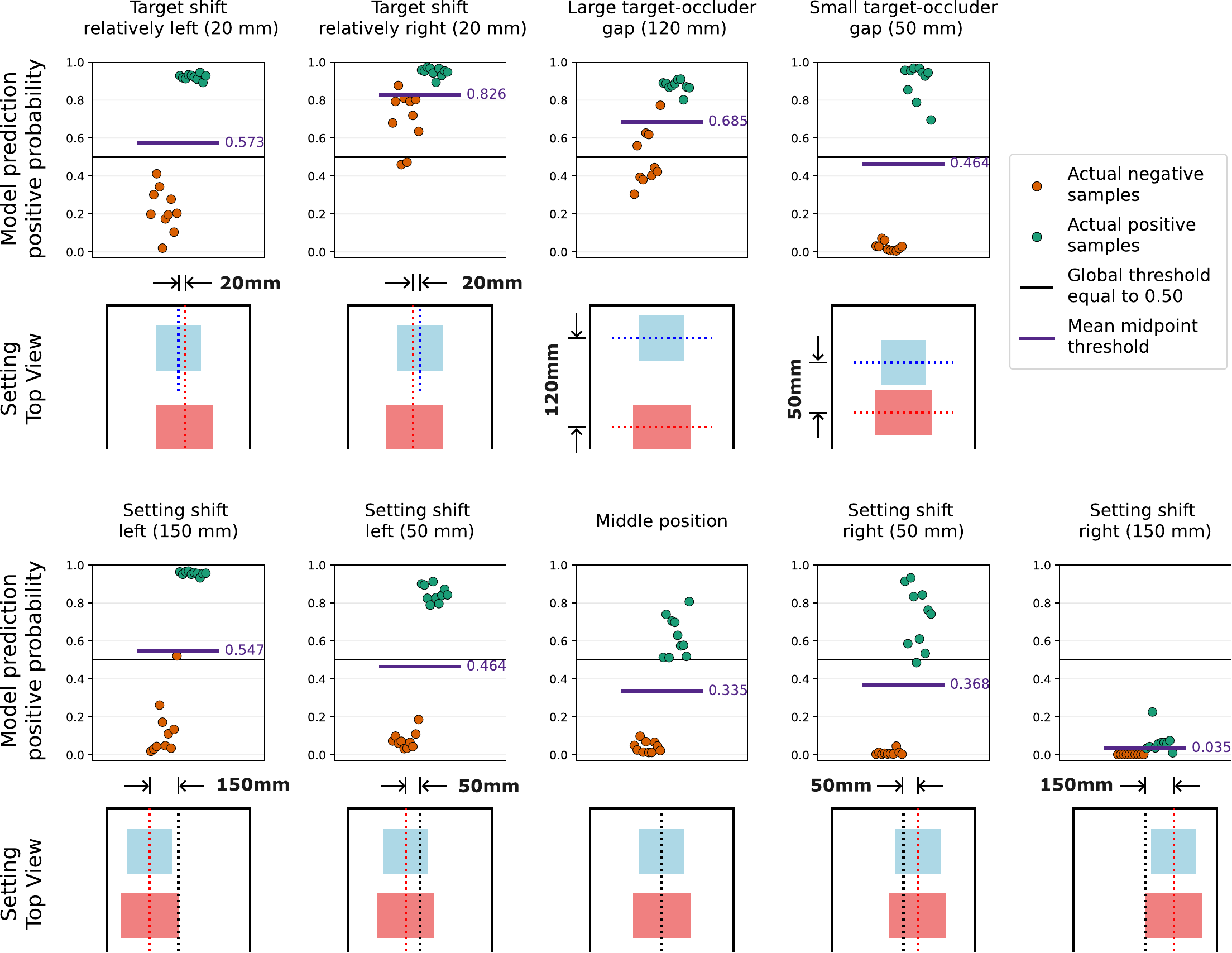}
\caption{Experimental validation across nine target--occluder geometry configurations, showing prediction scores and corresponding top-view.}
\label{fig:experimental_setup}
\end{figure}

A total of 5,700 synthetic samples, held out from the training dataset, and 180 experimental samples are used for validation. The synthetic samples are randomly selected from the generated dataset, while the experimental data are specifically arranged to evaluate model performance under spatial shifts and changes in the relative target–occluder position, including position offsets and different target–occluder gaps. The predicted probability of target presence from the model and corresponding top-view configurations are shown in Fig.~\ref{fig:experimental_setup}. Across the experimental configurations, positive samples generally receive higher prediction scores than negative samples. However, as the target--occluder setup shifts from left to right, the prediction scores change noticeably, with some right-shifted cases receiving lower positive scores. This indicates a position-dependent bias in the model. Changes in the relative target--occluder offset and separation also affect the score distribution, although the two classes remain well separated in most configurations.

\begin{table}[htbp]
\centering
\caption{Performance of the proposed model on synthetic and experimental data.}
\label{tab:result_tab}
\begin{tabular}{lcccccc}
\hline
Dataset & Accuracy & Precision & Recall & F1-score & AUC & BCE loss \\
\hline
Synthetic data ($n=5{,}700$) & 86.26\% & 86.28\% & 86.25\% & 86.26\% & 94.65\% & 0.3162 \\
Experiment data ($n=180$) & 86.67\% & 85.87\% & 87.78\% & 86.81\% & 91.88\% & 0.3894 \\
\hline
\end{tabular}
\end{table}

Table~\ref{tab:result_tab} summarizes the classification performance on both synthetic and experimental data. The synthetic test set reaches 86.26\% accuracy, 86.26\% F1-score, and 94.65\% AUC, while the experimental set achieves 86.67\% accuracy, 86.81\% F1-score, and 91.88\% AUC. The close precision and recall values on both datasets indicate balanced classification without a strong bias toward either class. Although the experimental BCE loss is slightly higher, the comparable overall performance demonstrates effective simulation-to-experiment transfer from synthetic training data to real SPAD LiDAR measurements.

\begin{figure}[htbp]
\centering
  \includegraphics[width=\linewidth]{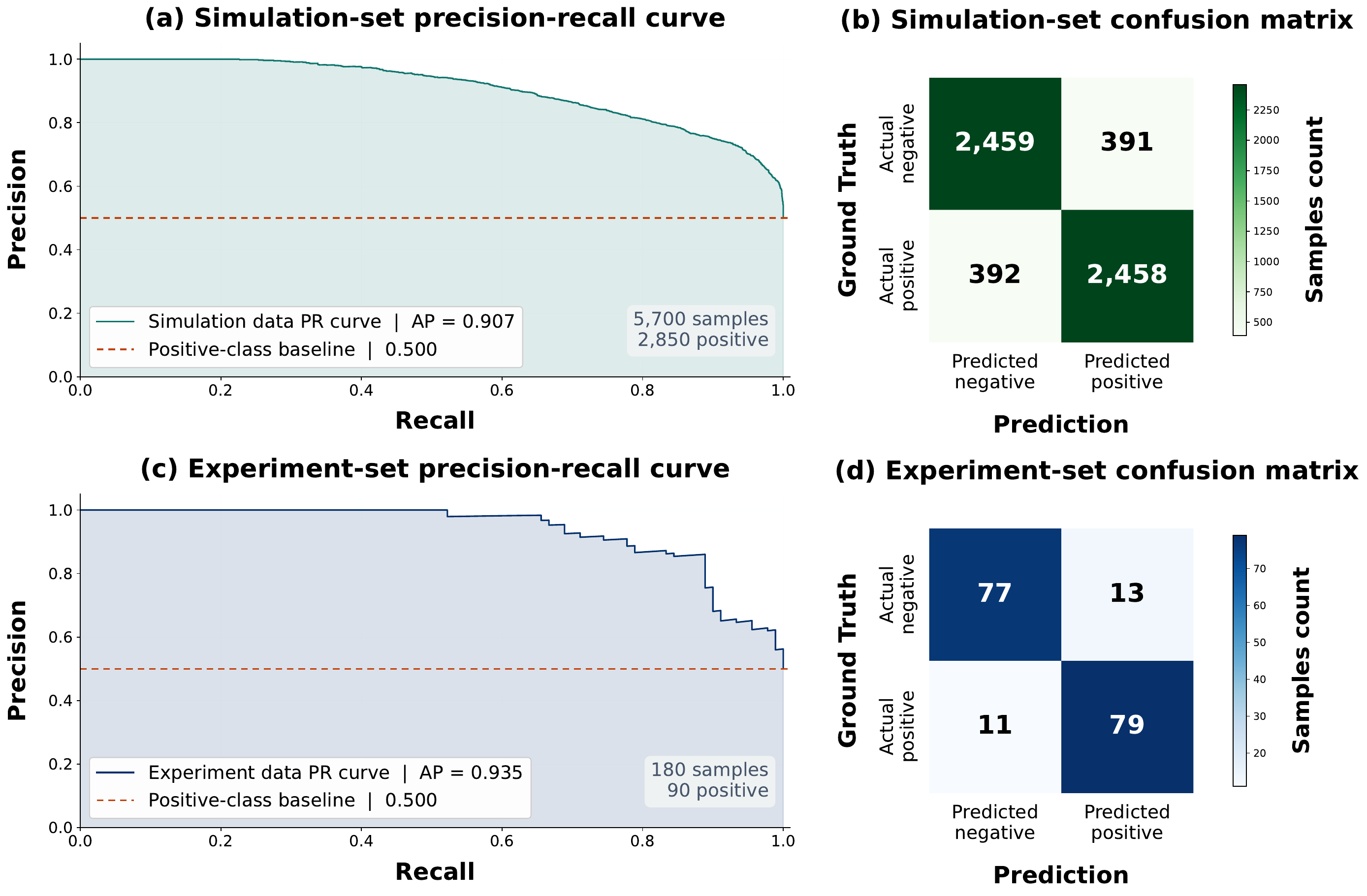}
\caption{Precision--recall curve and confusion matrix for simulation datasets (a,b) and experimental datasets (c,d).}
\label{fig:result_analysis}
\end{figure}

Further analysis is performed using precision--recall curves and confusion matrices, as shown in Fig.~\ref{fig:result_analysis}. On the synthetic test set, the model achieves an AP of 0.91 and an accuracy of 86.26\%, with 2458 true positives and 2459 true negatives. On the experimental set, the AP is 0.94 with an accuracy of 86.67\%, yielding 79 true positives and 77 true negatives. Although the experimental precision--recall curve is less smooth due to the smaller sample size, the similar accuracy and class balance indicate effective transfer from synthetic data to real SPAD LiDAR measurements. The slightly higher experimental AP should be interpreted cautiously because the experimental set contains only 180 samples and covers fewer scene configurations.

\section{Discussion}

The results show effective simulation-to-experiment transfer from synthetic training data to real measurements acquired with a low-cost commercial SPAD LiDAR. The model achieves similar performance on synthetic and experimental datasets, with AP values of 0.907 and 0.935 and accuracies of 86.26\% and 86.67\%, respectively. However, the experimental results reveal position-dependent bias, with prediction scores varying across target--occluder positions and separations, while sensor-specific pixel noise requires calibration to reduce the simulation-to-experiment gap. The relatively small experimental dataset and limited scene diversity should also be considered when interpreting the experimental performance. 

Overall, the results demonstrate the feasibility of simulation-only training for occlusion-based NLOS detection using low-cost commercial flash LiDAR. However, several limitations remain in the current dataset, such as limited range of geometries, occluder shapes, target positions, surface reflectivities, and sensor configurations. The model also remains sensitive to sensor-specific characteristics and spatial noise patterns, indicating that more accurate device modeling is needed to further reduce the simulation-to-experiment gap. Future work will therefore focus on increasing scene diversity, improving sensor modeling, and reducing spatial dependence to improve generalization to more complex NLOS environments. Beyond binary detection, the framework could also be extended to estimate target location, shape, and reflectivity, with the longer-term goal of reconstructing partially or fully occluded scenes using low-cost commercial flash LiDAR.

\section{Conclusion}

In this work, an occlusion-based NLOS scenario that differs from the conventional relay-based setting was studied. The problem is more ill-posed because there is no well-defined relay surface, and the measured transient contains complex multipath reflections. Our goal is to detect a target hidden behind and spatially separated from an occluder by identifying target-dependent patterns in transient histograms. Because no dataset exists for this setting, synthetic training samples were generated using a SYCL-accelerated Monte Carlo photon-path simulator. The simulator models multi-path propagation, while sensor-specific noise is added to reduce the simulation-to-experiment gap. The dataset includes variations in scene geometry and sensor conditions and is used to train a histogram-based classifier for hidden-target detection. The model is then validated on real measurements from a commercial TMF8829 SPAD flash LiDAR, achieving 86.67\% accuracy and an AP of 0.95 on 180 experimental samples. The source code for this work is available at \url{https://github.com/eelab-dev/LiDAR-Synth}, the model is available at \url{https://huggingface.co/eelab-dev/Occlusion_Based_NLOS_model}, and the dataset is available at \url{https://huggingface.co/datasets/eelab-dev/Occlusion_Based_NLOS}.

\section*{Acknowledgments}
The authors thank Prof. Jonathan Leach at Heriot-Watt University for his technical discussion on non-line-of-sight. Also, Mr Jamie Brown and the School of Engineering Mechanical Workshop for their assistance with the experimental setup, and the Royal Society (grant RGS\textbackslash R2\textbackslash 222007) for supporting the computing resources. Dr Istvan Gyongy is funded by UKRI's Quantum Enabled Position, Navigation and Timing (QEPNT) (grant EP/Z533178/1), UKRI Proof of Concept (UKRI6318), and EPSRC Photonics \& Quantum Accelerator UK Research and Innovation (PQA102). Dr Danial Chitnis is funded by UKRI's AI for Productive Research \& Innovation in eLectronics (APRIL) Hub (grant EP/Y029763/1).

\bibliography{sample}

\end{document}